# A minimal model of predator–swarm interactions

Yuxin Chen[1] and Theodore Kolokolnikov[2]

[1]Department of Engineering Sciences and Applied Mathematics, Northwestern University, Evanston, IL, USA
[2]Department of Mathematics and Statistics, Dalhousie University, Halifax, Canada

We propose a minimal model of predator–swarm interactions which captures many of the essential dynamics observed in nature. Different outcomes are observed depending on the predator strength. For a 'weak' predator, the swarm is able to escape the predator completely. As the strength is increased, the predator is able to catch up with the swarm as a whole, but the individual prey is able to escape by 'confusing' the predator: the prey forms a ring with the predator at the centre. For higher predator strength, complex chasing dynamics are observed which can become chaotic. For even higher strength, the predator is able to successfully capture the prey. Our model is simple enough to be amenable to a full mathematical analysis, which is used to predict the shape of the swarm as well as the resulting predator–prey dynamics as a function of model parameters. We show that, as the predator strength is increased, there is a transition (owing to a Hopf bifurcation) from confusion state to chasing dynamics, and we compute the threshold analytically. Our analysis indicates that the swarming behaviour is not helpful in avoiding the predator, suggesting that there are other reasons why the species may swarm. The complex shape of the swarm in our model during the chasing dynamics is similar to the shape of a flock of sheep avoiding a shepherd.

## 1. Introduction

Many species in nature form cohesive groups. Some of the more striking examples are schools of fish and flocks of birds, but various forms of collective behaviour occur at all levels of living organisms, from bacterial colonies to human cities. It has been postulated that swarming behaviour is an evolutionary adaptation that confers certain benefits on the individuals or group as a whole [1–5]. These benefits may include more efficient food gathering [6], predator avoidance in fish shoals [7] or zebra [4] and heat preservation in penguin huddles [8]. An example is the defensive tactics used by a zebra herd against hyaenas or lions [4]. These defence mechanisms may include evasive manoeuvres, confusing the predator, safety in numbers and increased vigilance [4,9,10]. On the other hand, a countervailing view is that swarming can also be detrimental to prey, as it makes it easier for the predator to spot and attack the group as a whole [1].

Figure 1 gives some idea of the variety and complexity of predator–swarm interactions that occur in nature. A common characteristic is the formation of empty space surrounding the 'predator' (or a human shepherd as in figure 1a). There is also a presence of a relatively sharp boundary of the swarm.

In this paper, we investigate a very simple particle-based model of predator–prey interactions which captures several distinct behaviours that are observed in nature. There are several well-known mechanisms whereby the prey tries to avoid the predator. One well-studied example is *predator confusion*, which occurs when the predator is 'confused' about which individual to pursue. Predator confusion decreases the predators' ability to hunt their prey. To quote Krause & Ruxton [3, p. 19], 'predator confusion effect describes the reduced attack-to-kill ratio experienced by a predator resulting from an inability to single out and attack individual prey'. Bazazi *et al.* [12] studied marching insects and demonstrated that their collective behaviour functions partly as

Royal Society **Publishing**

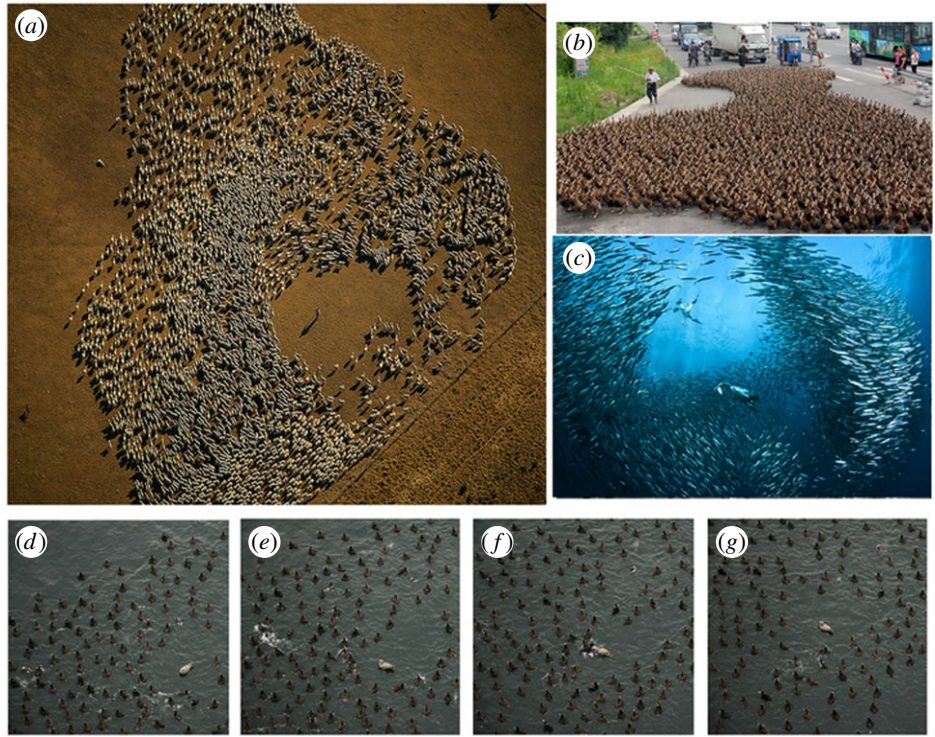

**Figure 1.** (*a*) Flock of sheep in Argentina avoiding the shepherd in the middle. (Photograph by Yann Arthus Bertrand, used with the permission of the author.) (*b*) A farmer walks 5000 ducks in Taizhou, China. (Source: BBC news.) (*c*) A baitball of sardines under attack by diving gannets. (Source: *The Telegraph*, Jason Heller/Barcroft Media.) (*d–g*) A flock of ducks in Vancouver, Canada, being pursued by a kleptoparasitic gull. Snapshots taken 2 s apart showing complex pursuit dynamics. (Photographs by Ryan Lukeman, used with the permission of the author. Photographs (*e,f*) also appear in [11].) (Online version in colour.)

an anti-predator strategy. During hunting, predators become confused when confronted with their prey swarm [13] and predator confusion was observed in 64% of the predator–prey systems studied in [14]. Predator–prey dynamics were also studied using computer models (e.g. [5,15–17]). Zheng *et al.* [15] studied a mathematical model of schools of fish, which demonstrates that collective evasion reduces the predator's success by confusing it. Olson *et al.* [16] used simulated coevolution of predators and prey to demonstrate that predator confusion gives a sufficient selective advantage for swarming prey. Similar preference for swarming in the presence of a confused predator was investigated in [5].

While there are many models in the literature that demonstrate complex predator–prey dynamics, most of these models are too complex to study except through numerical simulations. The goal of this paper is to present a minimal mathematical model which is carefully chosen so that (i) it is amenable to mathematical analysis and (ii) it captures the essential features of predator–prey interactions. A commonly used approach to swarm dynamics is to represent each prey by a particle that moves based on its interactions with other prey and its interaction with the predator. There is a large literature on particle models in biology, where they have been used to model biological aggregation in general [1,18–22] and locusts [21] or fish populations [15,23–27] in particular. This is the approach that we take in this paper as well.

We now introduce the model that we study in this paper. We assume that there are $N$ prey whose positions $x_j(t) \in \mathbb{R}^2$, $j = 1 \dots N$ follow Newton's law so that $m(\mathrm{d}^2/\mathrm{d}t^2)x_j + \mu(\mathrm{d}/\mathrm{d}t)x_j = F_{j,\mathrm{prey-prey}} + F_{j,\mathrm{prey-predator}}$. Here, $F_{j,\mathrm{prey-prey}} + F_{j,\mathrm{prey-predator}}$ is the total force acting on the $j$-th particle, $\mu$ is the strength of 'friction' force and $m$ is its mass. We make a further simplification that the mass $m$ is negligible compared with the friction force $\mu$. After rescaling to set $\mu = 1$, the model is then simply $(\mathrm{d}/\mathrm{d}t)x_j = F_{j,\mathrm{prey-prey}} + F_{j,\mathrm{prey-predator}}$, so that the prey moves in the direction of the total force. This reduces the second-order model to a first-order model, which makes it easier to analyse mathematically. Similar reduction was used, for example, in the analysis of locust populations [28] and other biological models [19,29]. Various forms can be considered for prey–prey interactions. To keep cohesiveness of the swarm, we consider the interactions which exhibit pairwise short-range repulsion and long-range attraction, averaged over all of the particles. For concreteness, we consider the endogenous prey–prey interaction of the form $F_{j,\mathrm{prey-prey}} = 1/N \sum_{k=1,k\neq j}^{N} (1/|x_j - x_k|^2 - a)(x_j - x_k)$. The term $x_j - x_k/|x_j - x_k|^2$ represents Newtonian-type short-range repulsion that acts in the direction from $x_j$ to $x_k$, whereas $-a(x_j - x_k)$ is a linear long-range attraction in the same direction. While more general attraction–repulsion dynamics can be considered, we concentrate on this specific form because more explicit results are possible. In particular, in the absence of exogenous prey–predator force, this particular interaction has been shown to result in uniform swarms [30,31]. In general, the distribution inside the swarm can vary and have fluctuations; however, uniform density of a swarm is often a good first-order approximation for many swarms. For example, Miller & Stephen [32] found that the flocks of sandhill cranes feeding in cultivated fields had distribution close to uniform, regardless of flock size. See [19, pp. 537–538], and references therein for further examples and discussion of prevalence of nearly uniform distribution of flocks in nature.

The prey–predator interactions are modelled in a similar fashion: again for concreteness assume that there is a single predator whose position we denote as $z(t) \in \mathbb{R}^2$. Assuming that the predator acts as a repulsive particle on the prey, we take $F_{j,\mathrm{prey-predator}} = b((x_j - z)/|x_j - z|^2)$, with $b$ being the strength of the repulsion. Finally, we model the predator–prey interactions as an attractive force in a similar way, $(d/dt)z = F_{predator-prey}$. We consider the simplest scenario where $F_{predator-prey}$ is the average over all predator–prey interactions and each individual interaction is a power law, which decays at large distances; the prey then moves in the direction of the average force. These assumptions result in the following system:

$$\frac{dx_j}{dt} = \frac{1}{N}\sum_{k=1,k\neq j}^{N}\left(\frac{x_j - x_k}{|x_j - x_k|^2} - a(x_j - x_k)\right) + b\frac{x_j - z}{|x_j - z|^2} \quad (1.1)$$

and

$$\frac{dz}{dt} = \frac{c}{N}\sum_{k=1}^{N}\frac{x_k - z}{|x_k - z|^p}. \quad (1.2)$$

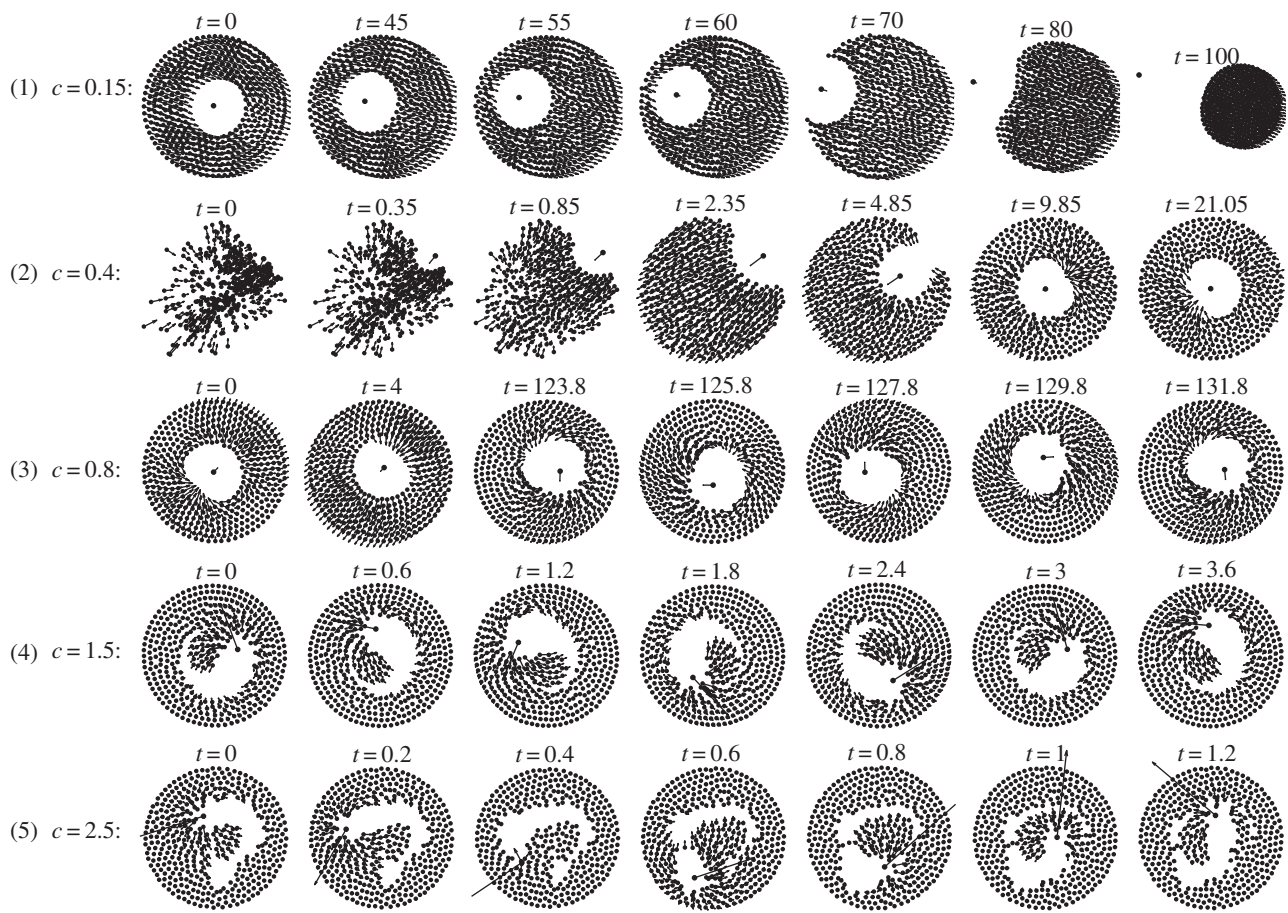

**Figure 2.** Predator–prey dynamics using the model (1.1) and (1.2). Parameters are $n = 400$; $a = 1$; $b = 0:2$; $p = 3$; and $c$ is given. The bifurcation values for $c$ are $c_0 = 0:2190$ and $c_{hopf} = 0:7557$ (see result 3.1). The velocity vector of the predator is also shown. First row: $c < c_0$; the swarm escapes completely. Second row: $c_0 < c < c_{hopf}$; predator catches up with the swarm but gets 'confused' and the swarm forms a stable ring around it. Third row: $c$ is just above $c_{hopf}$; regular oscillations are observed. Fourth row: $c$ is further increased leading to complex periodic patterns. Fifth row: the predator is able to 'catch' the prey (see §4); chaotic behaviour is observed.

To illustrate the results and motivate the analysis in this paper, consider the numerical simulations of the particle model (1.1) and (1.2) shown in figure 2. We use the strength $c$ of the predator–prey attraction as the control parameter, with other parameters as given in the figure. In the second row with $c = 0.4$, random initial conditions for prey and predator positions are taken inside a unit square. The swarm forms a 'ring' of constant density with a predator at the centre of the ring. Our first result is to fully characterize this ring in the limit of large swarms; see result 2.1. Our main result characterizes the stability of

this ring. In result 3.1, we show that the ring is stable whenever $2 < p < 4$ and

$$\frac{ba^{(2-p)/2}}{(1+b)^{(2-p)/2}} < c < \frac{a^{(2-p)/2}}{b^{(2-p)/2} - (1+b)^{(2-p)/2}}. \quad (1.3)$$

With parameters as chosen in figure 2 this corresponds to $0.2190 < c < 0.7557$. When $c$ is decreased below 0.2910 (row 1), the ring becomes unstable and the predator is 'expelled' out of the ring; the swarm escapes completely. A very different instability appears if $c$ is increased above 0.7557 (row 3). In this case, we show that the ring also becomes unstable *owing to the presence of oscillatory instabilities*, whereby the predator 'oscillates' around the 'centre' of the swarm. After some transients, the system settles into a 'rotating pattern' where the predator is continually chasing after its prey, without being able to fully catch up to it. As $c$ is further increased (row 4), the motion becomes progressively chaotic until the predator is finally able to catch the prey (row 5).

Our approach is to take the continuum-limit $N \to \infty$ of (1.1) and (1.2), which results in the non-local integro-differential equation model [19–22].

$$\rho_t(x, t) + \nabla \cdot (\rho(x, t)v(x, t)) = 0; \quad \int_{\mathbb{R}^2} \rho(y, t)dy = 1, \quad (1.4)$$

$$v(x, t) = \int_{\mathbb{R}^2}\left(\frac{x - y}{|x - y|^2} - a(x - y)\right)\rho(y, t)dy + b\frac{x - z}{|x - z|^2} \quad (1.5)$$

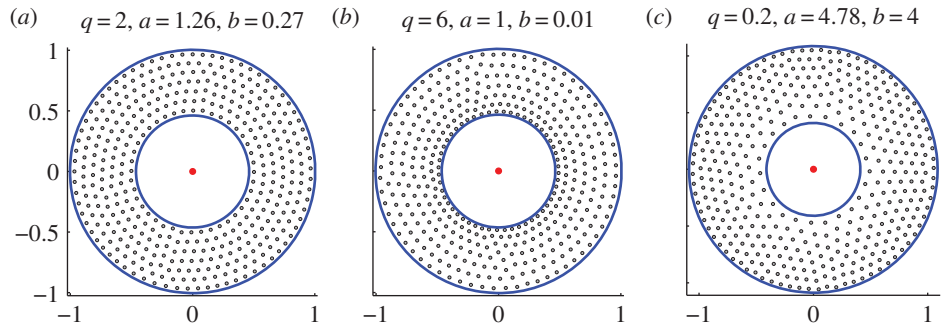

*(a)* $q = 2, a = 1.26, b = 0.27$   *(b)* $q = 6, a = 1, b = 0.01$   *(c)* $q = 0.2, a = 4.78, b = 4$

**Figure 3.** Steady state for model (4.2) and (1.2) for parameters as given and with $p = q$, $c = 10$. These states were computed by starting with random initial conditions and as such they appear to be stable. Solid circles correspond to the continuum-limit asymptotics (4.4). (*a*) $q = 2$, constant density swarm. (*b*) $q > 2$, swarm is denser towards the inner boundary. (*c*) $q < 2$, swarm is denser towards the outer boundary. (Online version in colour.)

and

$$\frac{\mathrm{d}z}{\mathrm{d}t} = c \int_{\mathbb{R}^2} \frac{y - z}{|y - z|^p} \rho(y, t) \mathrm{d}y. \tag{1.6}$$

Here, $\rho(x, t)$ denotes the density distribution of the prey swarm at position $x \in \mathbb{R}^2$ so that $\int_{\mathbb{R}^2} \rho(y, t) \mathrm{d}y = 1$ and $v(x, t)$ is the swarm's velocity field. The system (1.4)–(1.6) is obtained by choosing the initial density to be $\rho(x) = 1/N \sum_{j=1}^{N} \delta(x - x_j)$, where $\delta$ is the delta function. Equation (1.4) simply reflects the conservation of mass of the original prey system (1.1) (as no prey particles are created or destroyed); with the mass normalized so that $\rho(x, t)$ represents a probability distribution. By taking different pairwise endogenous forces, the steady state to (1.1) and (1.2) with no exogenous force ($b = 0$) presents a wide variety of patterns [33–35]. Similar equations have been used to model animal aggregation in [21,28,36–39]. The classical Keller–Segel model for chemotaxis also contains a Newtonian intra-species interaction [40,41]. Aggregation models also appear in material science [42–44], vortex motion [45–48] where Newtonian potential arises for vortex density evolution and granular flow [49,50].

We now summarize the paper. In §2, we construct the steady-state solution consisting of a ring of prey particles of uniform density that surround the predator at the centre. In §3, we study its stability. We conclude with some extensions of the model and discussion of some open problems in §4.

## 2. 'Confused' predator ring equilibrium state

We start by constructing the 'ring' steady state of the model (1.4)–(1.6), as shown in the last picture of the second row of figure 2. Consider a steady state for which the predator is at the centre of the swarm, surrounded by the prey particles. The predator is 'trapped' at the centre of the prey swarm while the prey forms a concentric annulus where the repulsion exerted by the predator cancels out owing to the symmetry. We state the main result as follows.

**Result 2.1.** *Define*

$$R_1 = \sqrt{\frac{b}{a}} \quad \text{and} \quad R_2 = \sqrt{\frac{1 + b}{a}}. \tag{2.1}$$

*The system* (1.4)–(1.6) *admits a steady state for which* $z = 0$, $\rho$ *is a positive constant inside an annulus* $R_1 < |x| < R_2$, *and is zero otherwise.*

Figure 3*a* illustrates this result. For parameters as shown in the figure, the discrete model (1.1) and (1.2) generates a stable ring steady state, which is shown with dots. Solid curves show the continuum result (2.1), in excellent agreement with the discrete model (1.1) and (1.2).

The fact that the density is constant inside a swarm is a result of the careful choice of the forces in (1.1): namely, the nonlinearities are both Newtonian. The proof of result 2.1 follows closely [30,31] and uses the method of characteristics, a common technique to find steady states in the aggregation model.

**Derivation of result 2.1.** Define the characteristic curves $X(X_0, t)$ which start from $X_0$ at $t = 0$

$$\frac{\mathrm{d}X}{\mathrm{d}t} = v(X, t); \quad X(X_0, 0) = X_0. \tag{2.2}$$

Using (1.4), along the characteristic curves $x = X(X_0, t)$, $\rho(x, t)$ satisfies

$$\frac{\mathrm{d}\rho}{\mathrm{d}t} = -(\nabla_x \cdot v)\rho. \tag{2.3}$$

Note that $\nabla_x \cdot ((x - z)/|x - z|^2) = \Delta_x (\ln |x - z|) = 2\pi\delta(x - z)$ so that from (1.5) we obtain

$$\nabla_x \cdot v = \int_{\mathbb{R}^2} [2\pi\delta(x - y) - 2a]\rho(y)\mathrm{d}y + 2\pi b\delta(x - z)$$
$$= 2\pi\rho(x) - 2aM, \quad x \neq z, \tag{2.4}$$

where $M = \int_{\mathbb{R}^2} \rho(y)\mathrm{d}y$ is conserved. Then (2.3) becomes

$$\frac{\mathrm{d}\rho}{\mathrm{d}t} = (2aM - 2\pi\rho)\rho, \tag{2.5}$$

which has a solution $\rho(X(X_0, t), t)$ approaching $aM/\pi$ as $t \to \infty$ and independent of the location, as long as $\rho(X_0, 0) > 0$.

Next, we seek a steady state such that $\rho$ is constant inside $\mathcal{A}$, $\rho$ zero outside $\mathcal{A}$, where $\mathcal{A}$ is an annulus $R_1 \leq |x| \leq R_2$, with $R_1$, $R_2$ and $z$ to be determined. Using the identity

$$\int_{|y| \leq R} \frac{x - y}{|x - y|^2} \mathrm{d}y = \begin{cases} \pi R^2 \dfrac{x}{|x|^2} & |x| > R, \\ \pi x, & |x| < R, \end{cases} \tag{2.6}$$

and for $x \in \mathcal{A}$, we compute

$$v(x) = \int_{\mathcal{A}} \left[ \frac{x - y}{|x - y|^2} - a(x - y) \right] \rho(y, t)\mathrm{d}y + \frac{b(x - z)}{|x - z|^2}$$
$$= \pi\rho x \left(1 - \frac{R_1^2}{|x|^2}\right) - ax\pi(R_2^2 - R_1^2)\rho + \frac{b(x - z)}{|x - z|^2}. \tag{2.7}$$

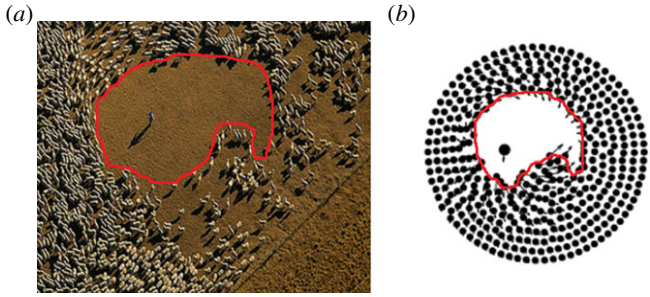

**Figure 4.** (*a*) The empty region surrounding the shepherd from figure 1*a* is shown with a curve. (*b*) Similar region observed in simulations of (1.1) and (1.2). (Online version in colour.)

The assumption of the steady state implies that (2.7) is zero for all $x \in \mathcal{A}$, which in turn implies that $z = 0$, $\pi - a\pi (R_2^2 - R_1^2) = 0$ and $\pi\rho(-R_1^2) + b = 0$ so that $R_1 = \sqrt{b/a}$ and $R_2 = \sqrt{(1+b)/a}$. Conversely, with this choice of $R_1$ and $R_2$, $v = 0$ whenever $\rho \neq 0$. Moreover by symmetry, $dz/dt = 0$ so that $v$, $\rho$, $z$ as in result 2.1 constitute a true steady state of (1.4)–(1.6).

## 3. Transition to chasing dynamics

As illustrated in figure 2, the ring steady-state configuration can transition to a moving configuration in two ways: if the predator strength $c$ is sufficiently decreased, the swarm will escape the predator. If $c$ is increased past another threshold, the predator becomes more 'focused' and less 'confused', resulting in 'chasing dynamics' which can lead to very complex periodic or chaotic behaviour. Similar dynamics can be observed in nature, as figure 4 illustrates. The onset of these dynamics can be understood as a transition from stability to an instability (i.e. bifurcation) of the ring steady state. The destabilizing perturbation corresponds to the translational motion of the predator as well as the inner or outer boundary of the ring.[1]

To understand these bifurcations, we consider the perturbations of the inner boundary, outer boundary, as well as the predator itself. These perturbations are of the form

$$\text{Inner boundary:} \quad x = R_1 e^{i\theta} + \varepsilon_1 e^{\lambda t}, \tag{3.1}$$

$$\text{Outer boundary:} \quad x = R_2 e^{i\theta} + \varepsilon_2 e^{\lambda t} \tag{3.2}$$

and

$$\text{Predator:} \quad z = 0 + \varepsilon_3 e^{\lambda t}, \tag{3.3}$$

where $\varepsilon_i \ll 1$. Note that this form of perturbation preserves the total mass which is an invariant of the model. In appendix A, we show that $\lambda$ satisfies the eigenvalue problem

$$(R_2^2 - R_1^2)\lambda \begin{pmatrix} \varepsilon_1 \\ \varepsilon_2 \\ \varepsilon_3 \end{pmatrix} = A \begin{pmatrix} \varepsilon_1 \\ \varepsilon_2 \\ \varepsilon_3 \end{pmatrix}, \tag{3.4}$$

where

$$A = \begin{pmatrix} -b-1 & b & 1 \\ -b - \dfrac{b}{1+b} & b & \dfrac{b}{1+b} \\ -c\left(\dfrac{b}{a}\right)^{(2-p)/2} & c\left(\dfrac{1+b}{a}\right)^{(2-p)/2} & c\left[\left(\dfrac{b}{a}\right)^{(2-p)/2} - \left(\dfrac{1+b}{a}\right)^{(2-p)/2}\right] \end{pmatrix}.$$

The eigenvalues of $A$ are given by $\lambda = 0$ and $\lambda = \lambda_\pm$ which satisfy $\lambda_\pm^2 + B\lambda_\pm + C = 0$, where $B = 1 - c\{(b/a)^{(2-p)/2} - ((1+b)/a)^{(2-p)/2}\}$ and $C = \frac{c((1+b)/a)^{(2-p)/2} - b}{1+b}$. The eigenvalues $\lambda_\pm$ are stable (i.e. $Re(\lambda_\pm) < 0$) if and only if $B > 0$ and $C > 0$. Note that, when $c = 0$, we get $B = 1$, $C < 0$ so that $\lambda_- < 0 < \lambda_+$ and the ring is unstable. As $c$ is increased, either $\lambda_+$ or $\lambda_-$ cross zero. This occurs precisely when $c = c_0$, where

$$c_0 = \frac{ba^{(2-p)/2}}{(1+b)^{(2-p)/2}}, \tag{3.5}$$

with $C > 0$ if and only if $c > c_0$. If $p \leq 2$, then $B > 0$ for all $c > c_0$ so that $Re(\lambda_\pm) < 0$. If $2 < p$, a Hopf bifurcation occurs when $B = 0$ with $C > 0$; i.e. when $c = c_{hopf} > c_0$, where

$$c_{hopf} = \frac{a^{(2-p)/2}}{b^{(2-p)/2} - (1+b)^{(2-p)/2}}. \tag{3.6}$$

Note $0 < c_0 < c_{hopf}$ if and only if $2 < p < 4$ (with $c_0 > c_{hopf}$ if $p > 4$, $c_{hopf} = \infty$ if $p = 2$ and $c_{hopf} < 0$ if $p < 2$). Therefore, $Re(\lambda_\pm) < 0$ if and only if one of the following holds: (i) $p \leq 2$ and $c > c_0$; (ii) $2 < p < 4$ and $c_0 < c < c_{hopf}$.

We summarize as follows.

**Result 3.1.** *Consider the ring steady state of (1.4)–(1.6) given in result 2.1. Let $c_0$, $c_{hopf}$ be as defined by (3.5) and (3.6), respectively.*

*The ring stability with respect to translational perturbations is characterized as follows:*

— *If $p \leq 2$: the ring is translationally stable if $c_0 < c$, and unstable if $c < c_0$.*
— *If $2 < p < 4$: the ring is translationally stable if $c_0 < c < c_{hopf}$. It is unstable owing to the presence of a negative real eigenvalue if $c < c_0$. As $c$ is increased past $c_{hopf}$, the ring is destabilized owing to a Hopf bifurcation.*
— *If $p > 4$: the ring is unstable for all positive $c$.*

This analysis reveals that there are three distinct regimes, which depend on the power exponent $p$ of the predator–prey attraction. If $p < 2$, then at close range the prey moves faster than the predator and can always escape. As a result, the predator can never catch the prey no matter how large $c$ is. The most interesting regime is $2 < p < 4$. As $c$ is increased just past $c_{hopf}$, complex periodic or chaotic chasing dynamics result, but the predator is still unable to catch the prey. The shape of the perturbation is reflected in the actual dynamics when $c$ is close to $c_{hopf}$ (such as in figure 2, row 3); however, as $c$ is further increased, nonlinear effects start to dominate and linear theory is insufficient to describe the resulting dynamics (see figure 2, rows 4 and 5). For even larger $c$, the predator finally 'catches' the prey; this is illustrated in figure 2, row 5; see §4 for further discussion of this.

Note that $c_0 = c_{hopf}$ when $p = 4$, in which case the stable band disappears. If $p > 4$, then $c_{hopf} < c_0$ and the ring configuration is unstable for *any* $c$. In this case, the swarm escapes completely if $c < c_{hopf}$ but chasing dynamics and catching of the prey can still be observed if $c > c_{hopf}$.

## 4. Discussion and extensions

The minimal model (1.1) and (1.2) supports a surprising variety of predator–swarm dynamics, including predator confusion, predator evasion and chasing dynamics (with rectilinear, periodic or chaotic motion).

Biologically, our model is useful in two ways. First, despite its simplicity our model has an uncanny ability to reproduce the complex shapes of a swarm in predator–prey systems. This is illustrated in figure 4. Second, the mathematical analysis of this model provides some rudimentary biological insight into general forces at play, which we now discuss.

Formula (3.6) shows that the prey–prey attraction that is responsible for prey aggregation, controlled by parameter $a$ in (1.1) and (1.2), is *detrimental* to prey: $c_{hopf}$ is a decreasing function of $a$ so that increasing $a$ makes it easier for the predator to catch the prey. This is also in agreement with several other studies. For example, Fertl and co-workers [51,52] observed groups of about 20–30 dolphins surrounding a school of fish and blowing bubbles underneath it in an apparent effort to keep the school from dispersing, while other members of the dolphin group swam through the resulting ball of fish to feed. In a survey [1], the authors suggest that factors other than predator avoidance, such as food gathering, ease of mating, energetic benefits or even constraints of physical environment, are responsible for prey aggregation. Our model also supports this conclusion.

The parameter $b$ in the model (1.1) and (1.2) can be thought of as the strength of prey–predator repulsion. Formula (3.6) shows that $c_{hopf}$ is an increasing function of $b$ so that increasing $b$ is beneficial to the prey.

The parameter $p$ can be vaguely interpreted as the predator 'sensitivity' when the prey is close to the predator and can be thought of as a measure of how sensitive the predator is to a nearby prey. Simple calculus shows that $c_{hopf}$ has a minimum at $p = p_{optimal}$ given by

$$p_{optimal} = 2 + 2 \frac{\ln\left(\ln\left((1+b)/a\right)\ln\left(b/a\right)\right)}{\ln\left((1+b)/b\right)}, \qquad (4.1)$$

provided that $b > a$ (no optimal $p$ exists otherwise with $c_{hopf} \to 0$ as $p \to \infty$; figure 5). From the point of view of the predator, this choice of sensitivity requires the least strength $c$ for success. It is unclear however whether this optimal value has a true biological significance or is simply an artefact of the model.

So far, we have concentrated on the onset of chasing dynamics as $c$ crosses $c_{hopf}$, as this value is computable analytically. This is a precursor to the predator catching the prey, but for values of $c$ just above $c_{hopf}$ the prey still escapes. Let us investigate further numerically what happens for larger values of $c$ when the predator can actually 'catch' the prey. For concreteness, we say that the prey is caught if the distance between it and the predator falls below a certain *kill radius*, which we take to be 0.01 in our simulations (numerically, the problem becomes unstable when this distance becomes too small as the velocity of the prey and predator increases without bound). Whenever the prey is caught, we remove it from the simulation (and decrease $N$ by 1 in (1.1) and (1.2)). Consider the parameters $p = 3$, $a = 1$, $b = 0.2$, $c = 1.8$ and suppose there are $N = 200$ prey initially. Figure 6 shows the number of prey as a function of time. It shows that the rate of consumption is higher with fewer individuals. The reader is invited to see the movie of these simulations.[2]

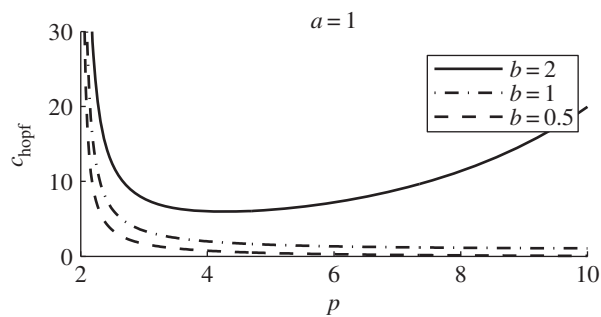

**Figure 5.** $c_{hopf}$ versus $p$ with $a = 1$ and $b$ as given. The curve has a minimum given by (4.1) if and only if $b > a$.

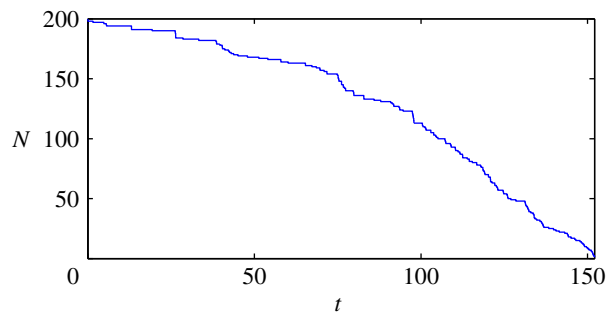

**Figure 6.** Number of prey remaining in a swarm during the hunt, as a function of time. Parameter values are $p = 3$, $a = 1$, $b = 0.2$ and $c = 1.8$. (Online version in colour.)

Let $c_{catch}$ be the smallest value of predator strength $c$ for which the predator is able to catch the prey. We compute this value using full numerical simulations of (1.1) and (1.2) for several values of $N$, while fixing the other parameters to be $p = 3$, $a = 1$, $b = 0.2$. The results are summarized in the following table:

| $N$ | 50 | 100 | 200 |
|---|---|---|---|
| $c_{catch}$ | 0.9 | 1.1 | 1.4. |

Note that $c_{catch}$ is increasing with $N$, which is also consistent with figure 6 showing that the kill rate increases when there are fewer particles. This suggests that all else being equal, having more individuals is beneficial to the prey, in that a higher predator strength $c$ is required to catch the prey when $N$ is increased. This may be owing to the fact that the predator becomes more 'confused' by the various individuals inside the swarm when there are more of them.

From a mathematical point of view, our analysis is rather non-standard: the main result is obtained by doing a stability analysis on the entire swarm in the continuum limit, which can be thought of as an infinite-dimensional dynamical system, or, alternatively, a non-local PDE–ODE system (1.4)–(1.6). Below we discuss several possible extensions of the model.

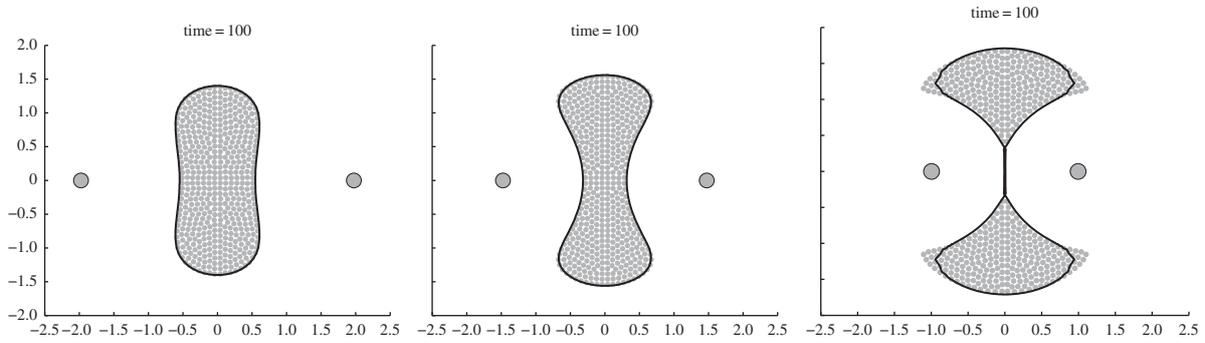

**Figure 7.** Symmetric steady states for (4.5) with $a = 1$, $b = 2$, $N = 500$, and with $M = 2$ predators located at $z_1 = (-d, 0)$ and $z_2 = (d, 0)$ with $d$ as given in the figure. Steady states are represented by the dots. The solid line is the boundary computed by using the continuum formulation (4.6). Note that, in the right-hand figure, the swarm separates into two groups.

## 4.1. Non-uniform state

The first extension is to replace the prey–predator interaction in (1.1) by a more general power nonlinearity, for example

$$\frac{dx_j}{dt} = \frac{1}{N} \sum_{k=1, k \neq j}^{N} \left( \frac{x_j - x_k}{|x_j - x_k|^2} - a(x_j - x_k) \right) + b \frac{x_j - z}{|x_j - z|^q}, \quad (4.2)$$

with the equation for the predator unchanged; the original model corresponds to $q = 2$. As before, there is a steady state with the predator $z = 0$ at the centre with the swarm forming a ring around it. Unlike the $q = 2$ case, the density of the swarm is no longer uniform. Using a computation similar to the $q = 2$ case, we find that, in the continuum limit, the density is given by

$$\rho(x) = \begin{cases} \dfrac{a}{\pi} - \dfrac{b(2-q)}{2\pi |x|^q} & \text{when } R_1 < |x| < R_2 \\ 0 & \text{otherwise,} \end{cases} \quad (4.3)$$

with $R_1$, $R_2$ satisfying

$$R_1 = \left( \frac{b}{a} \right)^{1/q}, \quad a(R_2^2 - R_1^2) - b/2(R_2^{2-q} - R_1^{2-q}) = 1, \quad (4.4)$$

result 2.1 is recovered by choosing $q = 2$ in (4.4).

From (4.3) we note that for $q < 2$, the density is higher further away from the predator; conversely for $q > 2$ the density is higher closer to the predator. This compares favourably with full numerical simulations as shown in figure 3. However, the computation of stability for the non-constant density state remains an open problem.

## 4.2. Multiple predators

It is easy to generalize (1.1) and (1.2) to include multiple predators. For example, replace (1.1) by

$$\frac{dx_j}{dt} = \frac{1}{N} \sum_{k=1, k \neq j}^{N} \left( \frac{x_j - x_k}{|x_j - x_k|^2} - a(x_j - x_k) \right) + \sum_{k=1}^{M} b \frac{x_j - z_k}{|x_j - z_k|^2}, \quad (4.5)$$

and replace $z$ by $z_k$ in (1.2) (more complex predator–predator interactions can similarly be added). Even more complex dynamics can be observed. Multi-species interaction has been studied in several other contexts recently, including crowd dynamics and pedestrian traffic [53,54], decision-making in the group with strong leaders [55] and generalization of the Keller–Segel model to multi-species in chemotaxis [56–58].

Here, we briefly consider the possible steady states of the swarm in the presence of two stationary predators (i.e. $c = 0$). Consider two predators located symmetrically at $z_1 = d$ and $z_2 = -d$. Figure 7 shows some of the possible steady states for various values of $d$. As $d$ is decreased, the swarm splits into two. The swarm is symmetric with respect to $x$- and $y$-axes but is not radially symmetric.

The solid curve in figure 7 shows the continuum limit of (4.5) which is obtained by computing the evolution of the boundary $\partial D$ of the swarm, while assuming that swarm density $\rho = 1/|D|$ is constant. Using the divergence theorem, the velocity can then be computed using only a one-dimensional integration

$$v(x) = \frac{1}{|D|} \int_{\partial D} \ln |x - y| \hat{n} \, dS(y) - ax + b \frac{x - z_1}{|x - z_1|^2} + b \frac{x - z_2}{|x - z_2|^2}, \quad (4.6)$$

where we assumed that the centre of mass of the swarm is at the origin, and where the area $|D| = \int_D dy = \frac{1}{2} \int_{\partial D} y \cdot \hat{n} \, dS(y)$ is also a one-dimensional computation.

## 4.3. Acceleration and other effects

Introducing acceleration allows for a more realistic motion. A more general model is

$$m_j \frac{d^2 x_j}{dt^2} + \mu_j \frac{dx_j}{dt} = \frac{1}{N} \sum_{\substack{k=1, \\ k \neq j}}^{N} F(|x_j - x_k|) \frac{x_j - x_k}{|x_j - x_k|} + G(|x_j - z|) \frac{x - z}{|x - z|} \quad (4.7)$$

and

$$M \frac{d^2 z}{dt^2} + \mu_0 \frac{dz}{dt} = \frac{1}{N} \sum_{k=1}^{N} H(|z - x_k|) \frac{z - x_k}{|z - x_k|}, \quad (4.8)$$

where $\mu_j$, $\mu_0$ are friction coefficients of prey and predator, respectively, and $m_j$, $M$ are their masses. Figure 8 illustrates some of the possible dynamics of these models. Even more complex models exist in the literature. For example, to obtain a more realistic motion for fish an alignment term is often included, which can lead to milling and flocking patterns even in the absence of predator [26,59,60].

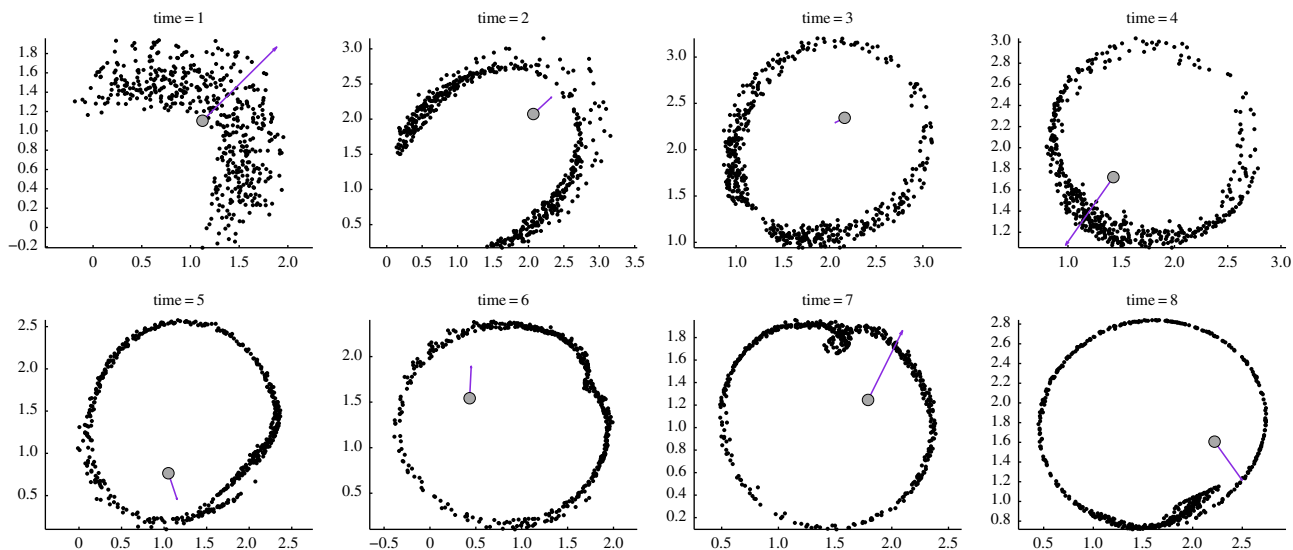

**Figure 8.** Predator–prey dynamics in second-order model (4.7) and (4.8) with $N = 500$, $F(r) = 1/r - r$, $G(r) = 1/r$, $H(r) = 1/(r^5 + 1)$ and $m_j$, $\mu_j$, $M$, $\mu_0 = 1$. (Online version in colour.)

Many models of collective animal behaviour found in the literature include terms such as zone of alignment, angle of vision, acceleration, etc. These terms may result in a more 'realistic-looking' motion, although it can be difficult in practice to actually measure precisely how 'realistic' it is (but see [11,61] for work in this direction). Moreover, the added complexity makes it very difficult to study the model except through numerical simulations. Our minimal model shows that these additional effects are not necessary to reproduce complex predator–prey interactions.

**Acknowledgements.** We are grateful to Ryan Lukeman and Yann Arthus Bertrand for providing us with their photographs. We also thank the anonymous referees for their valuable comments, which improved the paper significantly.

**Funding statement.** T.K. was supported by a grant from AARMS CRG in Dynamical Systems and NSERC grant no. 47050. Some of the research for this project was carried out while Y.C. and T.K. were supported by the California Research Training Program in Computational and Applied Mathematics (NSF grant DMS-1045536).

## Endnotes

[1] As discussed in the derivation of result 2.1, for large time, the density $\rho(x,t)$ rapidly approaches a constant on its support, $\rho \to aM/\pi$ and the equation for $\rho$ along characteristics is independent of the boundary shape or the form of predator–prey interactions (parameters $c$ and $p$ in (1.2)). As such, tracking the evolution of the boundary and the predator is sufficient to determine the stability of the ring state.
[2] We created a website which contains the movies showing the simulations of predator–swarm interactions from this paper. These can be viewed by following the link: http://goo.gl/BC6pyC.

## Appendix A

In this appendix, we derive eigenvalue problem (3.4) for the perturbations of the form (3.1)–(3.3). Let $o_i = \varepsilon_i e^{\lambda t}$. The velocity then becomes

$$v(x) = \rho \int_{B(o_2, R_2) \setminus B(o_1, R_1)} \left[ \frac{x - y}{|x - y|^2} - a(x - y) \right] dy + b \frac{x - z}{|x - z|^2}. \quad (A1)$$

Using (2.6) with $x \in B(o_2, R_2) \setminus B(o_1, R_1)$, we get

$$v(x) = \rho \left[ \pi x - \pi o_2 + a\pi x(R_1^2 - R_2^2) - \frac{\pi R_1^2(x - o_1)}{|x - o_1|^2} + a\pi(R_2^2 o_2 - R_1^2 o_1) \right] + \frac{b(x - o_3)}{|x - o_3|^2}. \quad (A2)$$

At the steady state $o_i = 0$ and $v = 0$ so that (A 2) simplifies to

$$v(x) = \rho\pi[(aR_2^2 - 1)o_2 - aR_1^2 o_1] + \rho\pi R_1^2 \left[ \frac{x}{|x|^2} - \frac{x - o_1}{|x - o_1|^2} \right] + b \left[ \frac{x - o_3}{|x - o_3|^2} - \frac{x}{|x|^2} \right]. \quad (A3)$$

On the inner boundary, we have $x = R_1 e^{i\theta} + \varepsilon_1 e^{\lambda t}$ and linearizing we obtain

$$v \sim \rho\pi[(aR_2^2 - 1)o_2 - aR_1^2 o_1] - \rho\pi o_1 e^{2i\theta} + \frac{bo_3 e^{2i\theta}}{R_1^2}.$$

Evaluating the perpendicular component $v_\perp = v \cdot e^{i\theta}$ yields

$$v_\perp \sim \left\{ \rho\pi[(aR_2^2 - 1)\varepsilon_2 - aR_1^2 \varepsilon_1] - \rho\pi\varepsilon_1 + \frac{b\varepsilon_3}{R_1^2} \right\} e^{\lambda t} \cos(\theta).$$

We equate $v = dx/dt = \lambda\varepsilon_1 e^{\lambda t}$ along the perpendicular component to finally obtain

$$\frac{\lambda}{\rho\pi}\varepsilon_1 = (-aR_1^2 - 1)\varepsilon_1 + (aR_2^2 - 1)\varepsilon_2 + \frac{b}{\rho\pi R_1^2}\varepsilon_3. \quad (A4)$$

The same computation along the outer boundary $x = R_2 e^{i\theta} + \varepsilon_2 e^{\lambda t}$ yields

$$\frac{\lambda}{\rho\pi}\varepsilon_2 = \left( -aR_1^2 - \frac{R_1^2}{R_2^2} \right)\varepsilon_1 + (aR_2^2 - 1)\varepsilon_2 + \frac{b}{\rho\pi R_2^2}\varepsilon_3. \quad (A5)$$

Next, we linearize predator equation (1.6) around ring steady state (2.1). We estimate

$$\int_{B(o_2,R_2)\setminus B(o_1,R_1)} \frac{x-o_3}{|x-o_3|^p}dx \sim \int_{B(o_2,R_2)\setminus B(o_1,R_1)} \frac{x}{|x|^p}dx$$
$$+ \int_{B(0,R_2)\setminus B(0,R_1)} \frac{px(x\cdot o_3) - o_3\,|x|^2}{|x|^{p+2}}dx$$
$$+ \text{h.o.t.},$$

where h.o.t. denotes higher order terms that are quadratic in $o_i$. We then compute explicitly

$$\int_{B(0,R_2)\setminus B(0,R_1)} \frac{px(x\cdot o_3) - o_3\,|x|^2}{|x|^{p+2}}dx = -\pi o_3(R_2^{2-p} - R_1^{2-p}),$$

and approximate

$$\int_{B(o_2,R_2)\setminus B(o_1,R_1)} \frac{x}{|x|^p}dx \sim \pi o_2 R_2^{2-p} - \pi o_1 R_1^{2-p}.$$

Linearizing predator equation (1.6) then yields

$$\frac{\lambda}{\rho\pi}\varepsilon_3 = -cR_1^{2-p}\varepsilon_1 + cR_2^{2-p}\varepsilon_2 + c(R_1^{2-p} - R_2^{2-p})\varepsilon_3. \quad (A\,6)$$

The three equations (A 6), (A 4) and (A 5) then yield a closed three-dimensional eigenvalue problem

$$\frac{\lambda}{\rho\pi}\begin{bmatrix}\varepsilon_1\\\varepsilon_2\\\varepsilon_3\end{bmatrix} = \begin{bmatrix} -aR_1^2-1 & aR_2^2-1 & \dfrac{b}{\rho\pi R_1^2} \\ -aR_1^2-\dfrac{R_1^2}{R_2^2} & aR_2^2-1 & \dfrac{b}{\rho\pi R_2^2} \\ -cR_1^{2-p} & cR_2^{2-p} & c(R_1^{2-p}-R_2^{2-p}) \end{bmatrix}\begin{bmatrix}\varepsilon_1\\\varepsilon_2\\\varepsilon_3\end{bmatrix}. \quad (A\,7)$$

Problem (3.4) is obtained by substituting (2.1) into (A 7).